\documentclass[prb,twocolumn,aps,superscriptaddress]{revtex4-1}

\usepackage{bm}
\usepackage[colorlinks=true,linkcolor=blue, citecolor=blue]{hyperref}
\usepackage{amsmath}
\usepackage{amssymb}
\usepackage{amsthm}

\usepackage{times}

\usepackage{microtype}
\usepackage{ccaption}
\usepackage{ragged2e}
  \captionnamefont{\bfseries \sffamily}
  \captiontitlefont{\small\sffamily}
  \captiondelim{ \textbar ~ }
  \captionstyle{\justifying}

\usepackage{xcolor}

\usepackage{amsfonts}
\usepackage{enumerate}
\usepackage{latexsym}
\usepackage{ifpdf}
\usepackage{graphicx}
\usepackage{color}
\usepackage{makeidx}
\usepackage{ulem}

\expandafter\ifx\csname package@font\endcsname\relax\else
\expandafter\expandafter
\expandafter\usepackage
\expandafter\expandafter
\expandafter{\csname package@font\endcsname}%
\fi
\begin{document}

\title{Artificial two-dimensional polar metal at room temperature}

\author{Yanwei Cao}
\email{yc874@physics.rutgers.edu}
\affiliation{Department of Physics and Astronomy, Rutgers University, Piscataway, New Jersey 08854, USA.}
\affiliation{CAS Key Laboratory of Magnetic Materials and Devices, Ningbo Institute of Materials Technology and Engineering, Chinese Academy of Sciences, Ningbo, Zhejiang 315201, China.}
\author{Zhen Wang}
\affiliation{Department of Physics and Astronomy, Louisiana State University, Baton Rouge, Louisiana 70803, USA.}
\affiliation{Department of Condensed Matter Physics and Materials Science, Brookhaven National Laboratory, Upton, New York 11973, USA.}
\author{Se Young Park}
\affiliation{Department of Physics, University of California Berkeley, Berkeley, California 94720, USA.}
\author{Y. Yuan}
\affiliation{Department of Materials Science and Engineering and Materials Research Institute, Pennsylvania State University, University Park, Pennsylvania 16802, USA.}
\author{Xiaoran Liu}
\affiliation{Department of Physics and Astronomy, Rutgers University, Piscataway, New Jersey 08854, USA.}
\author{S. Nikitin}
\affiliation{Department of Materials Science and Engineering and Materials Research Institute, Pennsylvania State University, University Park, Pennsylvania 16802, USA.}
\author{H. Akamatsu}
\affiliation{Department of Materials Science and Engineering and Materials Research Institute, Pennsylvania State University, University Park, Pennsylvania 16802, USA.}
\author{M. Kareev}
\affiliation{Department of Physics and Astronomy, Rutgers University, Piscataway, New Jersey 08854, USA.}
\author{S. Middey}
\affiliation{Department of Physics, University of Arkansas, Fayetteville, Arkansas 72701, USA.}
\affiliation{Department of Physics, Indian Institute of Science, Bangalore 560012, India}
\author{D. Meyers}
\affiliation{Department of Condensed Matter Physics and Materials Science, Brookhaven National Laboratory, Upton, New York 11973, USA.}
\author{P. Thompson}
\affiliation{XMas CRG, European Synchrotron Radiation Facility, Grenoble, Cedex 38043, France.}
\author{P. J. Ryan}
\affiliation{Advanced Photon Source, Argonne National Laboratory, Argonne, Illinois 60439, USA.}
\affiliation{School of Physical Sciences, Dublin City University, Dublin 9, Ireland.}
\author{P. Shafer}
\affiliation{Advanced Light Source, Lawrence Berkeley National Laboratory, Berkeley, California 94720, USA.}
\author{A. N'Diaye}
\affiliation{Advanced Light Source, Lawrence Berkeley National Laboratory, Berkeley, California 94720, USA.}
\author{E. Arenholz}
\affiliation{Advanced Light Source, Lawrence Berkeley National Laboratory, Berkeley, California 94720, USA.}
\author{V. Gopalan}
\affiliation{Department of Materials Science and Engineering and Materials Research Institute, Pennsylvania State University, University Park, Pennsylvania 16802, USA.}
\author{Yimei Zhu}
\affiliation{Department of Condensed Matter Physics and Materials Science, Brookhaven National Laboratory, Upton, New York 11973, USA.}
\author{Karin M. Rabe}
\affiliation{Department of Physics and Astronomy, Rutgers University, Piscataway, New Jersey 08854, USA.}
\author{J. Chakhalian}
\affiliation{Department of Physics and Astronomy, Rutgers University, Piscataway, New Jersey 08854, USA.}

\date{\today}

\begin{abstract}

\end{abstract}

\maketitle

\textbf{Polar metals, commonly defined by the coexistence of polar crystal structure and metallicity, are thought to be scarce because the long-range electrostatic fields favoring the polar structure are expected to be fully screened by the conduction electrons of  a metal. Moreover, reducing from three to two dimensions, it remains an open question whether a polar metal can exist. Here we report on the realization of a room temperature two-dimensional polar metal of the B-site type in tri-color (tri-layer) superlattices BaTiO$_3$/SrTiO$_3$/LaTiO$_3$. A combination of atomic resolution scanning transmission electron microscopy with electron energy loss spectroscopy, optical second harmonic generation, electrical transport, and first-principles calculations have revealed the microscopic mechanisms of periodic electric polarization, charge distribution, and orbital symmetry. Our results provide a route to creating all-oxide artificial non-centrosymmetric quasi-two-dimensional metals with exotic quantum states including coexisting ferroelectric, ferromagnetic, and superconducting phases.}

\begin{figure*}[htp]
\includegraphics[width=0.8\textwidth]{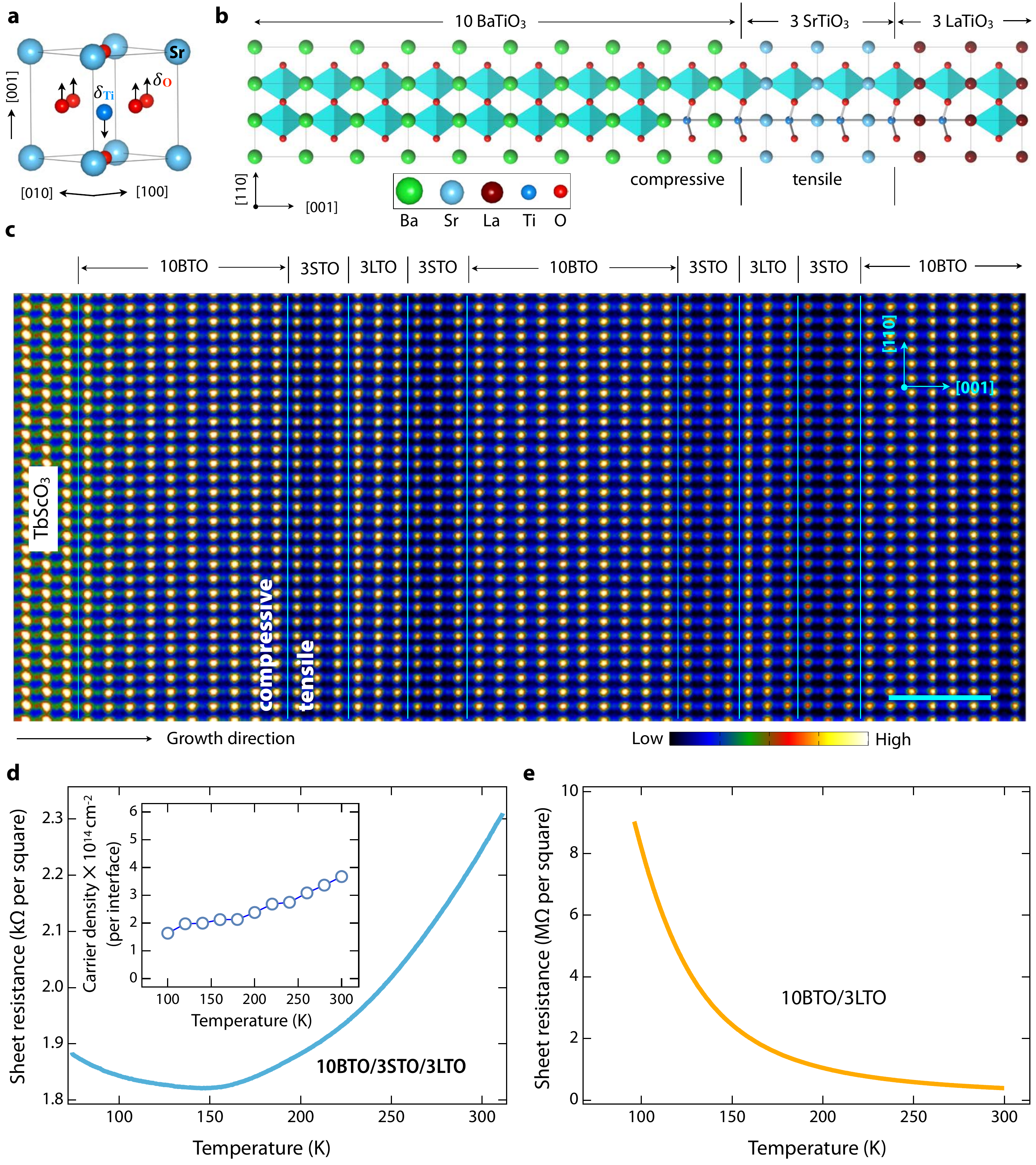}
\caption{\label{} Design and synthesis of BTO/STO/LTO superlattices. \textbf{a} Schematic polar displacements of Ti cations (\textit{$\delta$}$_{\textrm{Ti}}$, down arrow) and O anions (\textit{$\delta$}$_{\textrm{O}}$, up arrows) relative to the centrosymmetric plane of the corresponding unit cell. The cyan scale bar is 2 nm. \textbf{b} Sketch of tri-color heterostructure for designer two-dimensional polar metal. It is noted STO layer is under tensile strain ($\sim$ 1.3 \%) whereas BTO layer is under compressive strain ($\sim$ -1.1 \%). \textbf{c} HAADF-STEM image of 10BTO/3STO/3LTO on TbScO$_3$ substrate taken along [110] direction.  \textbf{d} Temperature-dependent sheet resistance (green curve) and carrier density (inset) of 10BTO/3STO/3LTO demonstrating a metallic behavior with large carrier density.  \textbf{e} Temperature-dependent sheet resistance of 10BTO/3LTO showing an insulating behavior. }
\end{figure*}

In modern materials science, the  design of new materials with emergent quantum ground states relies on the combination of structure-function and structure-composition relationships \cite{PRL-Anderson-1965,JMCC-Ben-2016,NM-Lu-2016,N-Lee-2005,N-Ya-2016,N-Mundy-2016,ARMR-2007-Schlom,NM-Shi-2013,Nature-Kim-2016,PRL-Ko-2010,SR-Fuj-2015}. The emergence of polar metals \cite{PRL-Anderson-1965,NM-Shi-2013,Nature-Kim-2016,PRL-Ko-2010,SR-Fuj-2015,JMCC-Ben-2016} is one of such  non-trivial  and   counter-intuitive examples  motivated by the search for  unconventional pairing in non-centrosymmetric  superconductors and  topology  protected spin currents.  Surprisingly, three extensive materials surveys \cite{CM-Hala-1998,JMCC-Ben-2016,NC-DP-2014} have revealed that oxide compounds are particularly scarce as polar metals. Based on the type of atomic displacements, polar metals with perovskite structure fall into two main categories \cite{JMCC-Ben-2016} - A-site  (e.g. positional shifts of Li, Nd, and Ca ions in LiOsO$_3$ \cite{NM-Shi-2013}, NdNiO$_3$ \cite{Nature-Kim-2016}, and CaTiO$_{3-\textit{$\delta$}}$ \cite{JMCC-Ben-2016}, respectively) or B-site dominated (e.g. shift of Ti ions in BaTiO$_{3-\textit{$\delta$}}$ \cite{JMCC-Ben-2016,PRL-Ko-2010,SR-Fuj-2015}) types. For the former category, recent theoretical work \cite{JMCC-Ben-2016} has suggested the absence of a fundamental incompatibility between the polarity and metallicity, whereas for the latter, polar displacements show a rapid decrease with increasing carrier concentration \cite{JMCC-Ben-2016,PRL-Ko-2010,SR-Fuj-2015}. Moreover,  since thin films  naturally hold a great potential for applications and   discovery of new physical phenomena \cite{Science-Ahn-2004}, creating polar metals via atomically-thin layering is a  promising approach. The challenge is that both free carrier screening and reduced dimensionality have been observed generally to suppress ferroelectric instabilities in perovskite oxides \cite{PRL-Wang-2012}.

\begin{figure*}[htp]
\includegraphics[width=0.8\textwidth]{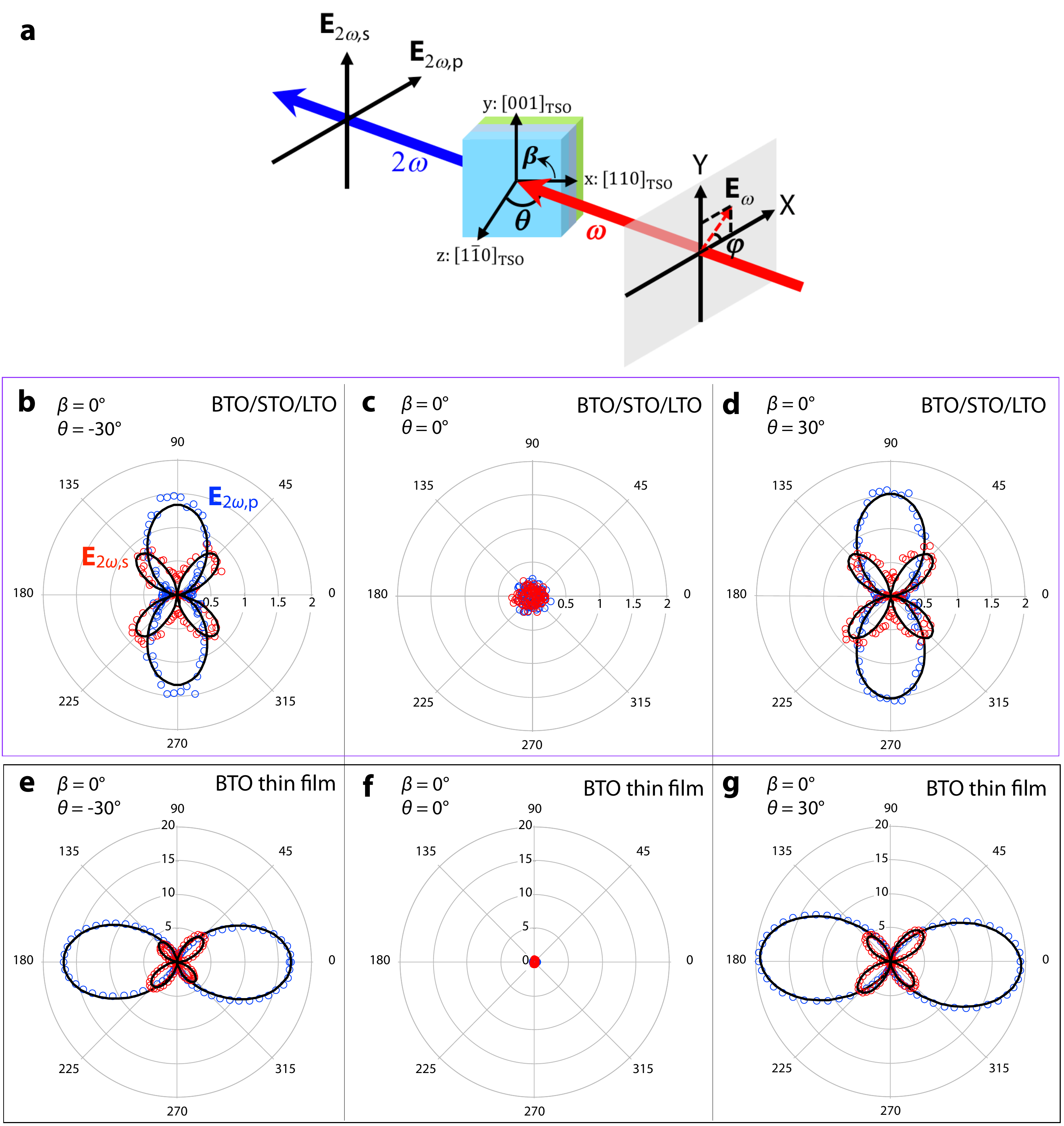}
\caption{\label{} SHG polarimetry measurement. \textbf{a} Schematics of the far-field transmission SHG setup with crystallographic directions of TbScO$_3$ substrate (subscript TSO) under sample coordinates (x, y, z). \textbf{E}$_{2\textit{$\omega$},\textrm{p}}$ (blue circles) and \textbf{E}$_{2\textit{$\omega$},\textrm{s}}$ (red circles) SHG signal on \textbf{b-d} BTO/STO/LTO and \textbf{c-g} BTO thin film (10 unit cells) measured by rotating the incident polarization \textit{$\varphi$} under three different incident angles \textit{$\theta$} = -30$^\circ$, $0^\circ$, $30^\circ$. Data with \textit{$\beta$} = 0$^\circ$ are plotted here (See Supplementary Fig. 3 a-f for \textit{$\beta$} = 90$^\circ$). Theoretical modeling of SHG signal (black solid lines) indicates 4$mm$ point group symmetry for BTO/STO/LTO with effective nonlinear optical coefficients \textbf{d}$_{33}$/\textbf{d}$_{15}$ $\approx$ -13.9, \textbf{d}$_{31}$/\textbf{d}$_{15}$ $\approx$ 1.3. In comparison, SHG signal on BTO thin film exhibits $mm$2 point group symmetry with \textbf{d}$_{33}$/\textbf{d}$_{15}$ $\approx$ 5.2, \textbf{d}$_{31}$/\textbf{d}$_{15}$ $\approx$ 0.1.}
\end{figure*}

\begin{figure*}[htp]
\includegraphics[width=0.8\textwidth]{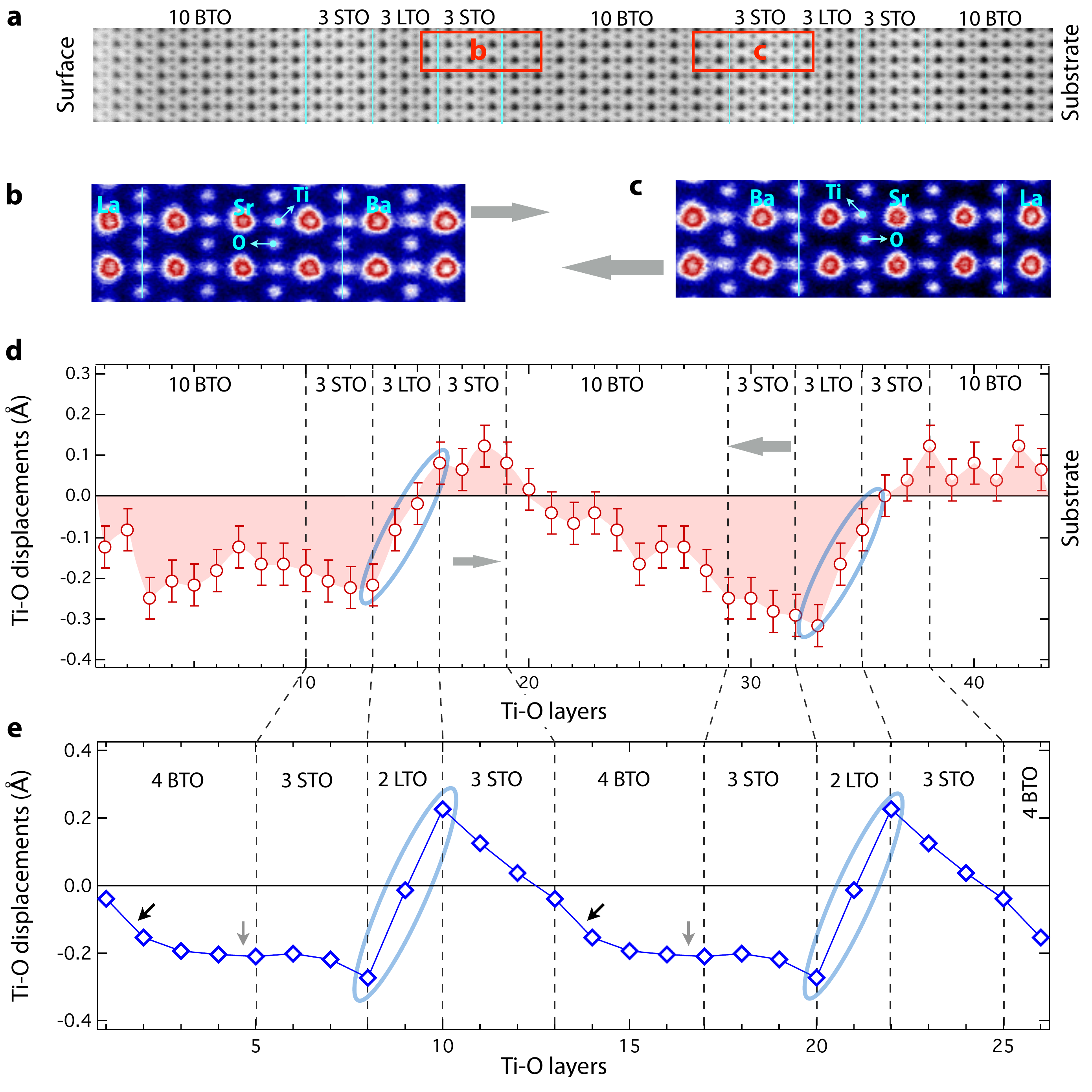}
\caption{\label{} Polar displacements. \textbf{a} ABF-STEM image of 10BTO/3STO/3LTO along [110] direction. \textbf{b,c} Enlarged images of red rectangular areas in \textbf{a} showing atomic positions of Ba, Sr, La, Ti, and O ions across interfaces. Gray arrows (in panel b-d) indicate the reversal directions of Ti-O polar displacements. \textbf{d} Experimentally layer-resolved Ti-O polar displacements in a 10BTO/3STO/3LTO superlattice. Here the Ti-O polar displacements (or $\Delta_{\textrm{Ti-O}}$) are defined as oxygen displacement with respect to Ti along the out-of-plane. The error bar shows the standard deviations of the averaged measurements for each vertical atomic layer. \textbf{e} Theoretically layer-resolved Ti-O polar displacements in a 4BTO/3STO/2LTO superlattice. Black and gray arrows indicate negative and zero slopes of the Ti-O displacements, respectively, in BTO layers, whereas blue ellipses highlight positive slopes in LTO layers.}
\end{figure*}

\begin{figure*}[htp!]
\includegraphics[width=1\textwidth]{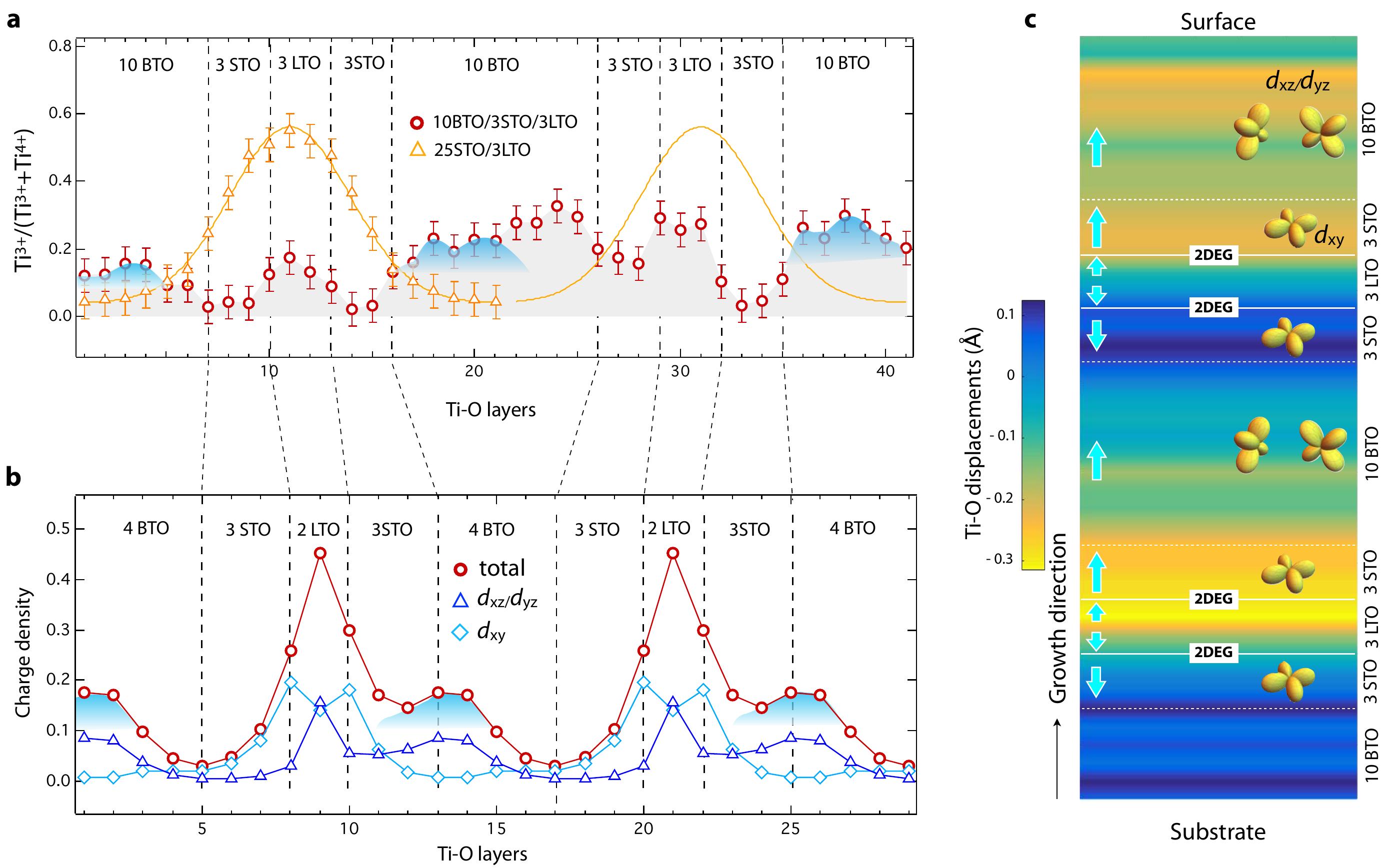}
\caption{\label{} Modulation of electric polarization, charge distribution, and orbital symmetry. \textbf{a} Experimentally layer-resolved charge distribution of 10BTO/3STO/3LTO (red circles) and reference sample 25STO/3LTO (yellow triangles) measured by STEM/EELS. It is noted the yellow solid lines are guide lines for clarity. Here the error bar indicates reasonable spacial resolution of STEM, the value of which is a constant $\pm$ 0.05 \AA. Blue shadows (in panel a and b) indicate the anomalous charge distribution in BTO layers. \textbf{b} Calculated orbitally resolved charge distribution for each TiO$_{2}$ layer in 4BTO/3STO/2LTO. Red circles indicate total projected density of states, whereas blue triangles and light blue rhombuses label the contribution from \textit{d}$_\textrm{xz}$/\textit{d}$_\textrm{yz}$ and \textit{d}$_\textrm{xy}$ orbitals, respectively. \textbf{c} Illustration of coexisting electric polarization and two-dimensional electron gas with theoretically proposed in-plane \textit{d}$_\textrm{xy}$-state as the preferential occupation. The image is plotted based on experimental Ti-O polar displacements (Fig. 3d). White solid lines schematically indicate the positions of two-dimensional electron gases, whereas cyan arrows mark the direction switching of Ti-O displacements. The strength and direction of polarization are labeled by the color map from blue to yellow.}
\end{figure*}

In the following, we design tri-color titanate heterostructures made of a layered arrangement of the ferroelectric alkaline-earth titanate BaTiO$_{3}$ (BTO), the paraelectric alkaline-earth titanate SrTiO$_{3 }$ (STO), and the Mott insulator rare-earth titanate LaTiO$_{3 }$ (LTO). In this design there are two inequivalent interfaces, BTO/STO and STO/LTO. The design idea is to transfer electrons from LTO into the STO layers forming two-dimensional electron gas (2DEG) at the interfaces \cite{Nature-Ohtomo-2002,PRL-Cao-2016}, which have a shared polar structure due to the presence of ferroelectric BTO \cite{Science-Tenne-2006,APL-Neaton-2003,PRB-Johnston-2005,PRL-Rag-2016}. In this paper, we present experimental measurements and first-principles calculations to show that a two-dimensional polar metal is thus realized in this tri-color structure with coexisting polar displacements of Ti and O sublattices as well as metallicity in TiO$_2$ atomic layers at the 2DEG interfaces. 

\textbf{Two-dimensional electron gas.}
Figure 1 shows the emergence of 2DEG at the STO/LTO interfaces of BTO/STO/LTO. Experimentally, ultra-thin tri-color titanate superlattices consisting of (BTO)$_{10}$/(STO$)_3$/(LTO)$_3$  (where the subscript refers to  the number of unit cells) as well as reference samples of (BTO)$_{10}$/(LTO)$_3$ superlattice and BTO thin film were synthesized on TbScO$_3$ (110) single crystal substrates by pulsed layer deposition in a layer-by-layer mode (See Fig. 1a,b, Supplementary Fig. 1, and Methods for more details). High crystallinity of the layers and good epitaxy were confirmed  by in-situ reflection-high-energy-electron-diffraction (RHEED)\ (Supplementary Fig. 1). To determine the atomic scale structure of the samples, the interfacial structure and composition of BTO/STO/LTO were investigated by cross-sectional scanning transmission electron microscopy (STEM) with electron energy loss spectroscopy (EELS). Figure 1c shows a high-angle annular dark-field (HAADF) STEM image of the tri-color superlattice, revealing high-quality continuous and coherent interfaces without phase separation. In the \textit{Z}-contrast HAADF image, the expected layer thickness and  designed sequence of three layers -...LTO/STO/BTO/STO/LTO...- are clearly distinguishable by the different intensities due to the difference in scattering power of the layers. Additionally, as shown in Supplementary Fig. 1d,  the periodicity of the growth sequence was further confirmed by the periodic variation of the out-of-plane lattice parameters of individual BTO, STO, and LTO layers. In addition, to eliminate concerns about possible chemical inter-diffusion at the interfaces, EELS spectroscopic imaging was performed at the Sc L$_{2,3}$-, Ti L$_{2,3}$-, Ba M$_{4,5}$- and La M$_{4,5}$-edges (Supplementary Fig. 2). As engineered with interfacial charge transfer, low temperature electrical transport measurements of BTO/STO/LTO verified the expected metallicity and large carrier density of conduction electrons in all tri-color samples ($n_\textrm{c}\sim$ 10$^{14}$ cm$^{-2}$, see Fig. 1 d,e). In the metallic BTO/STO/LTO, due to insulating BTO/STO (3\textit{d}$^{0}$ band) and BTO/LTO (Fig. 1e) interfaces the contribution of metallicity in BTO/STO/LTO is from the 2DEG of STO/LTO interfaces \cite{Nature-Ohtomo-2002,PRL-Cao-2016}. In sharp contrast to itinerant electrons at STO/LTO interfaces, it is noted the electrons at BTO/LTO interfaces are localized, the microcosmic mechanism of which needs further study.

\begin{figure*}[htp]
\includegraphics[width=0.8\textwidth]{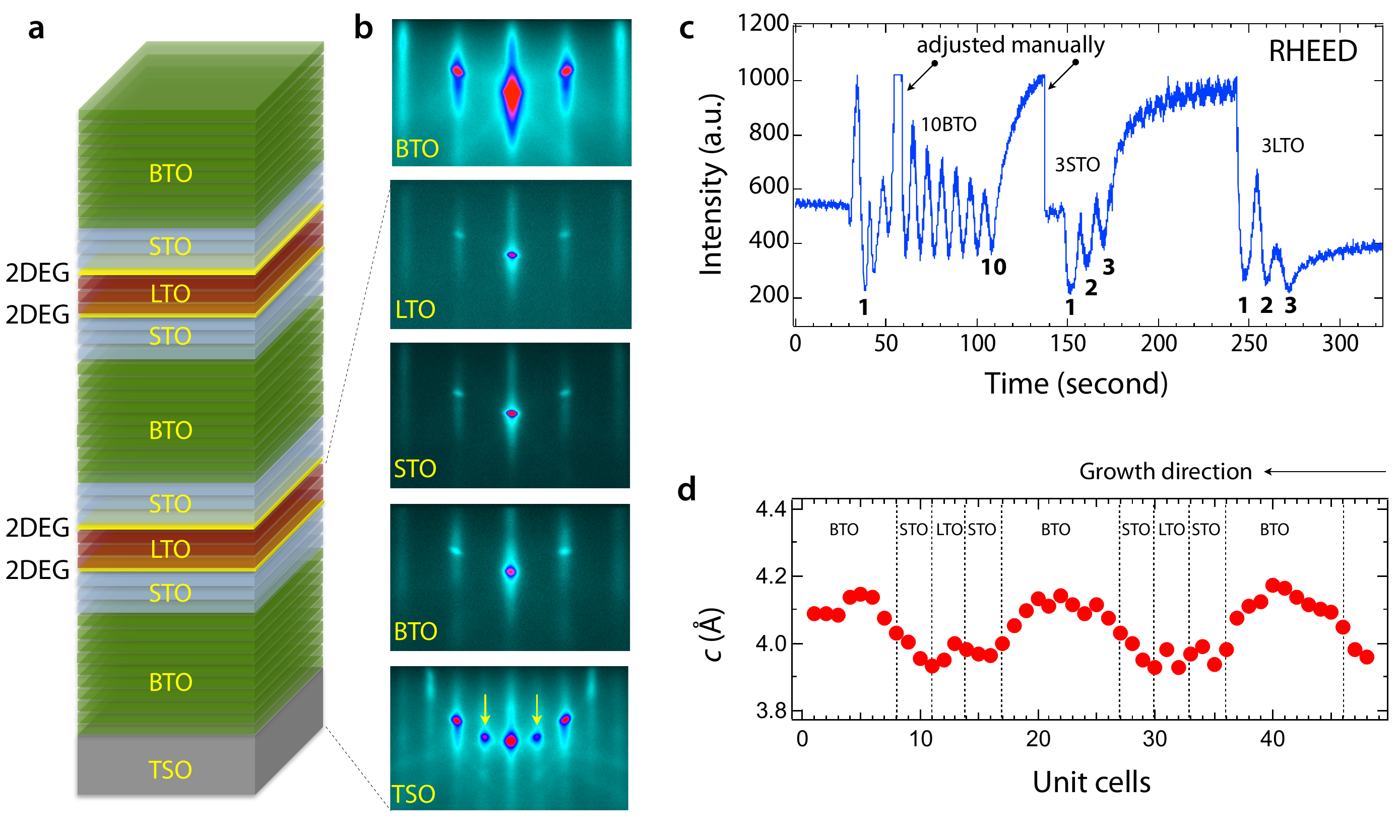}
\caption{\textbf{Supplementary Figure 1} Sample synthesis and characterization. \textbf{a}~~~Block sketch of growth sequences of BaTiO$_3$/SrTiO$_3$/LaTiO$_3$ superlattice, in which two-dimensional electron gases (2DEG) are formed at SrTiO$_3$/LaTiO$_3$ interfaces. \textbf{b}~~~RHEED patterns for TbScO$_3$ (TSO) substrate, BaTiO$_3$ (BTO), SrTiO$_3$ (STO), LaTiO$_3$ (LTO), and the top BaTiO$_3$ (BTO) layers, respectively, during the growth of (BTO)$_{10}$/(STO)$_3$/(STO)$_3$ (where the subscript refers to the number of unit cells). Yellow arrows indicate half-order-peaks of the orthorhombic structure of TSO substrate. \textbf{c}~~~Partial RHEED intensity oscillation during growth of (BTO)$_{10}$/(STO)$_3$/(STO)$_3$. \textbf{d}~~~Layer-dependent out-of-plane lattice parameter \textit{c} measured from STEM image, demonstrating the periodic synthesis of BTO, STO, and LTO layers of (BTO)$_{10}$/(STO)$_3$/(LTO)$_3$.}
\end{figure*}

\textbf{Polar distortions of crystal structures.}
To characterize the presence of polar displacements in BTO/STO/LTO,  we carried out optical SHG measurement, using a far-field transmission geometry as shown in Fig. 2a. Systematic polarimetry measurements on BTO/STO/LTO superlattice and BTO thin film were performed by measuring \textrm{p}- and \textrm{s}-polarized SHG signals \textbf{E}$_{2\textit{$\omega$},\textrm{p}}$ and \textbf{E}$_{2\textit{$\omega$},\textrm{s}}$ while rotating the incident polarization \textit{$\varphi$} in two different orientations \textit{$\beta$} = 0$^\circ$ and $90^\circ$ for three incident angles \textit{$\theta$} = -30$^\circ$, $0^\circ$, $30^\circ$ (see Supplementary Fig. 3 a-f for polarimetry with  \textit{$\beta$} = 90$^\circ$). As seen in Fig. 2b-g, both BTO/STO/LTO superlattice and the BTO thin film show a clear SHG signal, indicating they are both polar structures. 
The decreased maximum SHG intensity in BTO/STO/LTO superlattice as compared to BTO thin film is consistent with the results shown further on that indicate opposite out-of-plane polarizations between several layers and sometimes within one layer of the superlattice, which will lead to a partial phase cancellation of the SHG response from different layers of the superlattice.  In addition, remarkable differences in the SHG polarimetry were observed between BTO/STO/LTO superlattice and the BTO thin film. Specifically, the \textrm{p}-polarized SHG signal \textbf{E}$_{2\textit{$\omega$},\textrm{p}}$ of BTO/STO/LTO superlattice (Fig. 2b-d) has a maxima at incident polarizations of \textit{$\varphi$} = 90$^\circ$ and $270^\circ$, whereas the BTO thin film (Fig. 2e-g) shows its maximum at incident polarizations of \textit{$\varphi$} = 0$^\circ$ and $180^\circ$. A detailed theoretical modeling of SHG polarimetry revealed a net 4$mm$ point group symmetry with an out-of-plane polarization for BTO/STO/LTO superlattice grown on orthorhombic TbScO$_3$ (110) substrates, with effective nonlinear optical coefficients \textbf{d}$_{33}$/\textbf{d}$_{15}$ $\approx$ -13.9 and \textbf{d}$_{31}$/\textbf{d}$_{15}$ $\approx$ 1.3. In sharp contrast, the BTO thin film grown on the same substrate  shows a net $mm$2 point group symmetry with an out-of-plane polarization, with \textbf{d}$_{33}$/\textbf{d}$_{15}$ $\approx$ 5.2 and \textbf{d}$_{31}$/\textbf{d}$_{15}$ $\approx$ 0.1. As a reference, we also performed SHG polarimetry on the $c$-domain of a bulk BaTiO$_3$ single crystal (See Supplementary Fig. 3 g-i), which indicates a $4mm$  symmetry and  \textbf{d}$_{33}$/\textbf{d}$_{15}$ $\approx$ 3.3  and  \textbf{d}$_{31}$/\textbf{d}$_{15}$ $\approx$ 0.3, which are close to the values obtained from the BTO thin film and very different from BTO/STO/LTO superlattice. The breaking of the 4-fold to a 2-fold rotation symmetry in the 10 unit cells thick BTO thin film likely reflects the slight substrate anisotropy; this symmetry lowering is absent in the average $4mm$ symmetry of the thicker superlattice structure where this subtle substrate influence appears to have diminished. Ratios of nonlinear coefficients indicate intrinsic material properties as against microstructural effects; the dramatic changes in the ratios of between the BTO film and BTO/STO/LTO superlattice, strongly suggest that in addition to the BTO layers, the STO/LTO layers contribute a qualitatively distinct SHG behavior in the superlattice structure. 

\begin{figure*}[htp]
\includegraphics[width=0.7\textwidth]{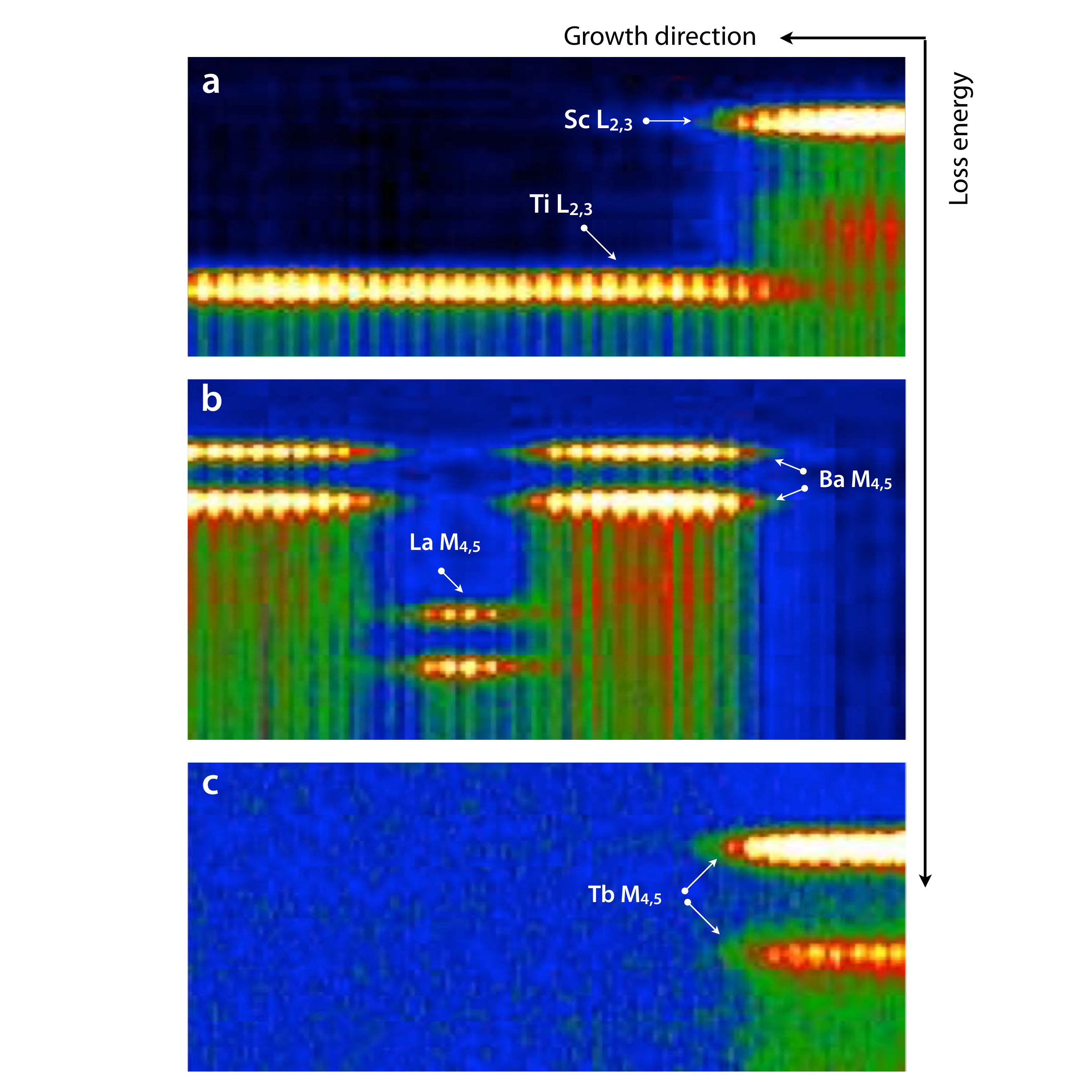}
\caption{\textbf{Supplementary Figure 2} EELS spectra maps of 10BTO/3STO/3LTO. \textbf{a}~~~Ti L$_{2,3}$- and Sc L$_{2,3}$-edges spectra. \textbf{b}~~~Ba M$_{4,5}$- and La M$_{4,5}$-edges. \textbf{c}~~~Tb M$_{4,5}$-edge.}
\end{figure*}

\textbf{Polar displacements at the atomic scale.}
Having established the presence of polar displacements averaged over the overall tri-color structure by SHG, next we investigate  the microscopic details of centrosymmetry breaking of TiO$_6$ octahedra in an atomic layer resolved way. To address this, high-resolution HAADF- and ABF-STEM imaging were carried out, which allows  to directly observe and extract the precise atomic positions of all constituent atoms, including oxygen across the interfaces. As shown in Fig. 3a-c, significant Ti-O polar displacements, i.e. relative shifts of Ti and O along the out-of-plane direction, are observed in the BTO/STO/LTO tri-color structure, which is consistent with the inversion symmetry breaking revealed by SHG. Additionally, to determine  amplitudes and directions of the polar displacements  a detailed quantitative analysis of the ABF-STEM image (Fig. 3a) was  performed. Figure 3d shows the evolution of Ti-O polar displacements layer-by-layer across the interfaces. The Ti-O polar displacements are found to be as large as 0.3 \AA, that is almost 8 \% enlargement of the lattice parameters ($\sim$ 4 \AA~for bulk BTO, STO and LTO). Moreover, these large Ti-O  polar displacements not only exist  in BTO but also extend deep into the STO and  LTO layers. This behavior agrees well with SHG data and further corroborates the presence of the polarization in the STO layer. A striking feature is the periodic reversal of polar directions across the LTO layers. We attribute this to atomic displacements driven by local up-down symmetry breaking , typical of perovskite surfaces, at the STO/LTO interface.  This non-switchable polar distortion propagates into the other layers: note, for instance, that the directions of  polar displacements in the TiO$_2$-atomic layers labeled by  the gray  arrows in Fig. 3b-d point in opposite ways. To understand the microscopic origin of  the polar distortions observed in the STEM and SHG measurements, first-principles GGA + \textit{U} calculation were performed. Figure 3e shows the calculated Ti-O polar displacements from ground-state atomic structure. The two dominant features, broken inversion symmetry from the net negative Ti-O displacements manifested in the BTO layers and the inversion of Ti-O polar displacements across LTO layers, are in a good agreement with the experimental data (Fig. 3d). The broken inversion symmetry is primarily driven by the polar distortion in BTO  ($\sim$ 0.2 \AA), and clearly absent in  the symmetric STO/LTO superlattice (see Supplementary Fig. 4). These polar distortions propagate deeper in to STO layers, consistent with previous predictions for BTO/STO superlattices \cite{Science-Tenne-2006,APL-Neaton-2003,PRB-Johnston-2005}. Near the STO/LTO interface, the Ti-O distortions are predominantly affected by the ionic screening of positively charged (LaO)$^{1+}$ layers in which the Ti (O) ions move away (to) the center of LaO layers. This behavior results in the reversal of the Ti-O displacements across the LTO region \cite{PRL-Okamoto-2006}.

\begin{figure*}[htp]
\includegraphics[width=0.8\textwidth]{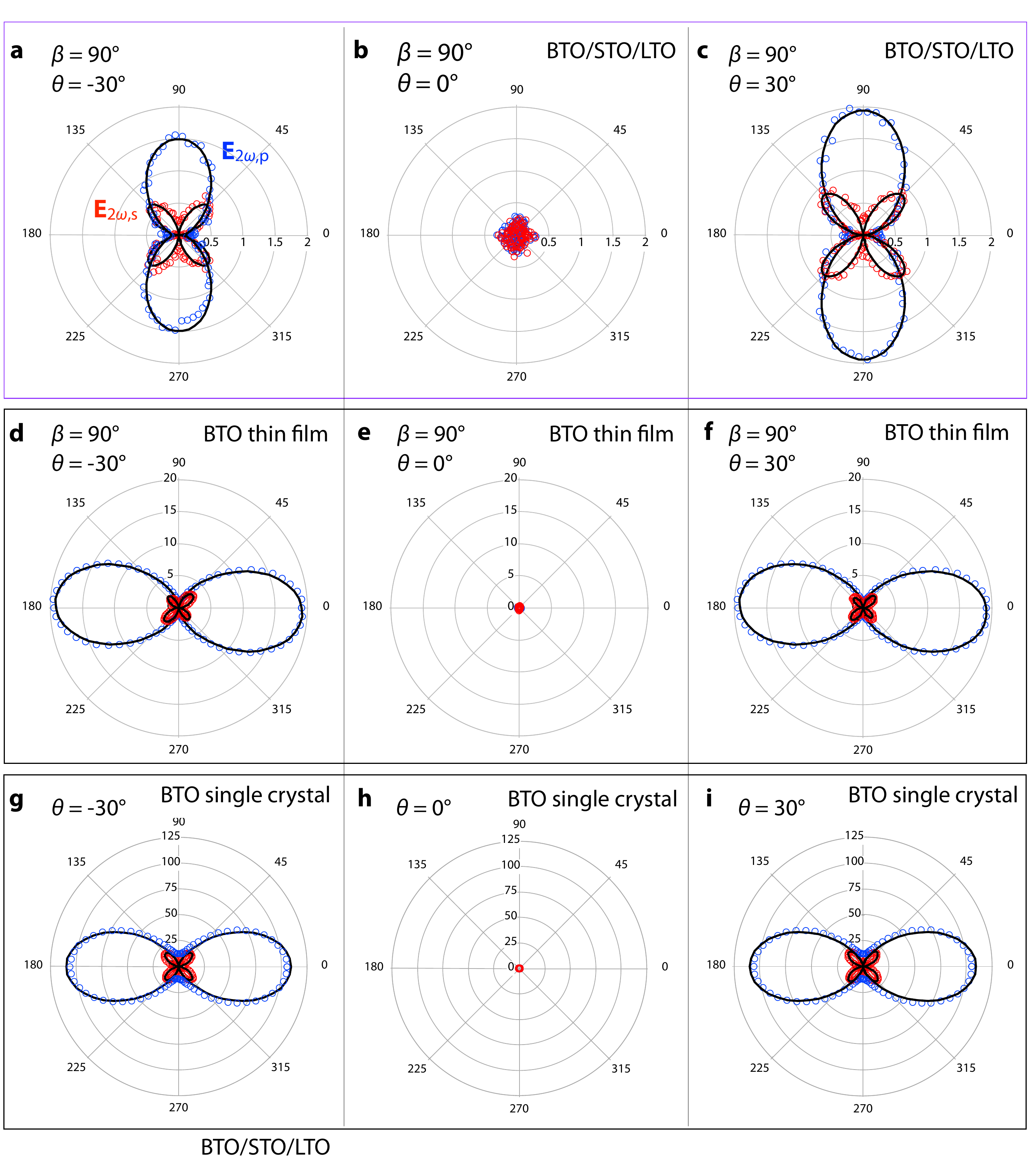}
\caption{\textbf{Supplementary Figure 3} SHG polarimetry measurement. \textbf{E}$_{2\textit{$\omega$},\textrm{p}}$ (blue circles) and \textbf{E}$_{2\textit{$\omega$},\textrm{s}}$ (red circles) SHG signal on \textbf{a-c}~~~10BTO/3STO/3LTO, \textbf{d-f}~~~10BTO thin film, and \textbf{g-i}~~~BTO single crystal measured by rotating the incident polarization \textit{$\varphi$} under three different incident angles \textit{$\theta$} = -30$^\circ$, $0^\circ$, $30^\circ$. Data with \textit{$\beta$} = 90$^\circ$ are plotted here. Theoretical modeling of SHG signal (black lines) indicates 4$mm$ point group symmetry for 10BTO/3STO/3LTO with effective nonlinear optical coefficients \textbf{d}$_{33}$/\textbf{d}$_{15}$ $\approx$ -13.9, \textbf{d}$_{31}$/\textbf{d}$_{15}$ $\approx$ 1.3. In comparison, SHG signal on 10BTO thin film and BTO single crystal exhibit $mm$2 point group symmetry with \textbf{d}$_{33}$/\textbf{d}$_{15}$ $\approx$ 5.2, \textbf{d}$_{31}$/\textbf{d}$_{15}$ $\approx$ 0.1 and \textbf{d}$_{33}$/\textbf{d}$_{15}$ $\approx$ 3.3, \textbf{d}$_{31}$/\textbf{d}$_{15}$ $\approx$ 0.3, respectively.}
\end{figure*}

\begin{figure*}[htp]
\includegraphics[width=0.8\textwidth]{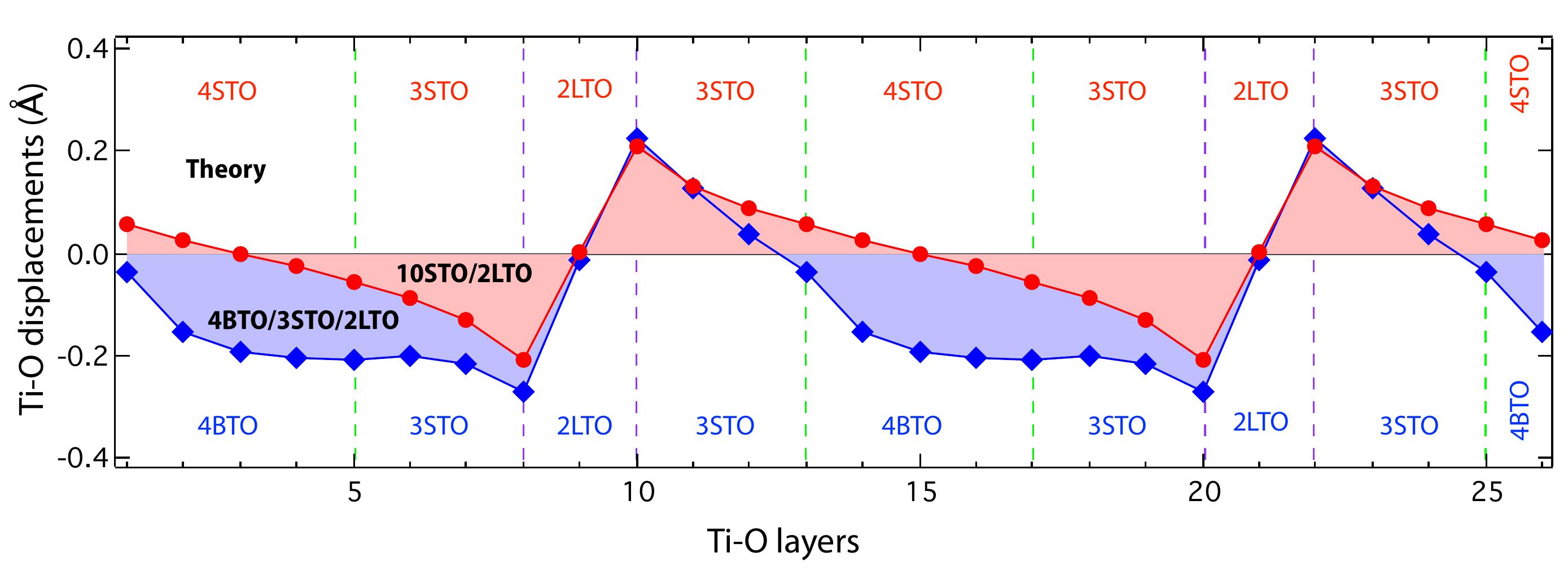}
\caption{\textbf{Supplementary Figure 4} Calculated Ti-O polar displacements. The comparison of polar displacements between tri-color 4BTO/3STO/2LTO (blue diamonds) and reference sample 10STO/2LTO (red dots) indicates the LTO layers plays an important role for the reversal of Ti-O polar displacements whereas the net polarization in 4BTO/3STO/2LTO is driven by polar distortions in BTO layers.}
\end{figure*}

\begin{figure*}[htp]
\includegraphics[width=1\textwidth]{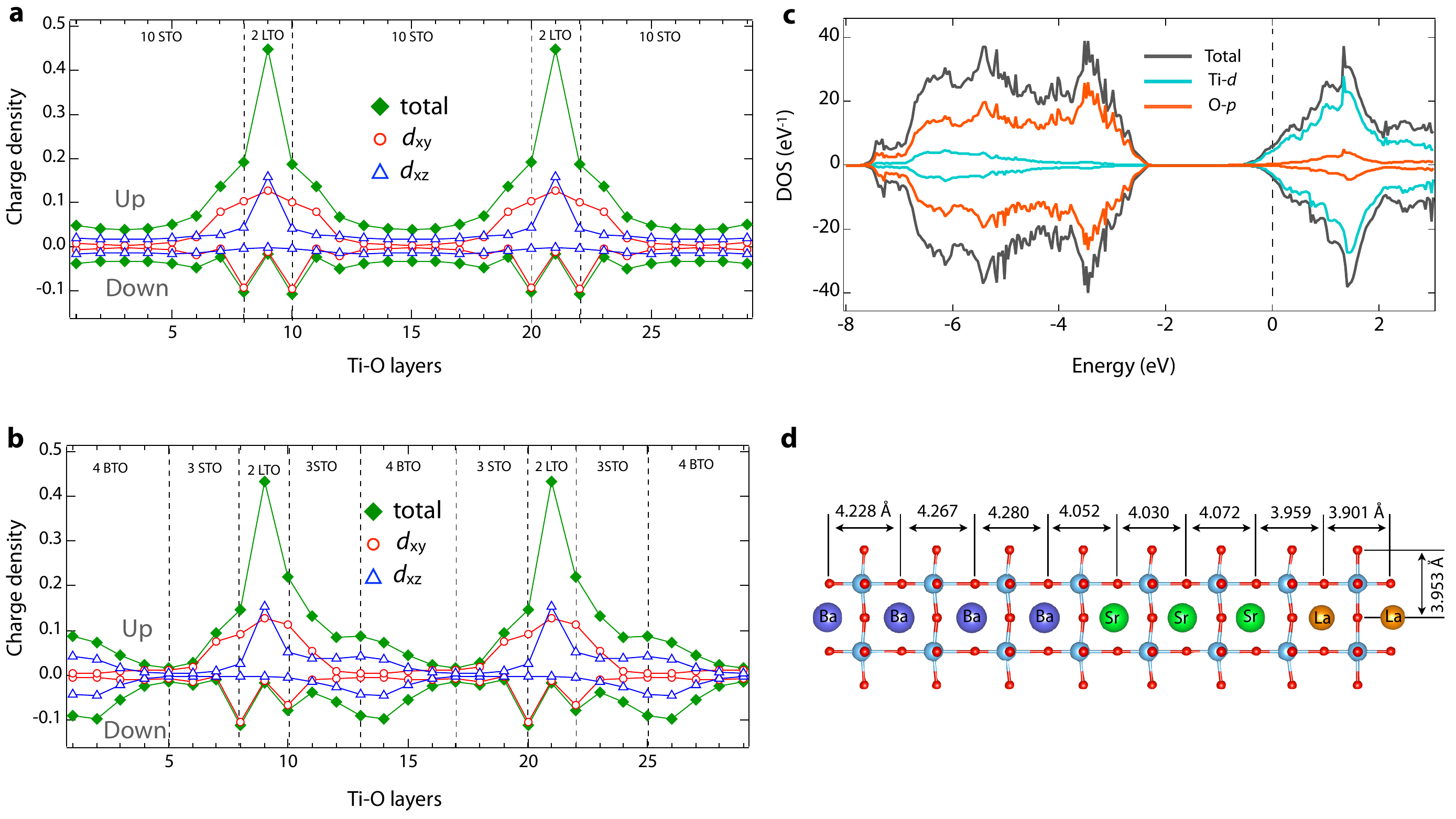}
\caption{\textbf{Supplementary Figure 5} Calculated orbitally resolved charge distribution for each TiO$_{2}$ layer. \textbf{a}~~~10STO/2LTO. \textbf{b}~~~4BTO/3STO/2LTO. \textbf{c}~~~Orbitally resolved projected density of state of 4BTO/3STO/2LTO superlattice. \textbf{d}~~~Calculated lattice parameters of TiO$_6$ octahedra across 4BTO/3STO/2LTO interfaces.}
\end{figure*}

\textbf{Modulation of interfacial metallicity and polarity.}
Since in complex oxides containing TM ions with open shells the spin, charge and orbital degrees of freedom are intimately coupled, next we discuss  the effect of the carrier modulation on  the orbital structure of  \textit{d}-electrons. Experimentally, the interfacial distribution of charge was  probed with atomic resolution STEM/EELS. In order to determine the valence state of Ti, we performed  fittings of the Ti L$_{2,3}$-edge EELS spectra across the interfaces by using reference spectra of Ti$^{4+}$ from bulk BTO and STO and Ti$^{3+}$ from bulk LTO.  Figure 4a shows the distribution of Ti$^{3+}$ (3\textit{d}$^1$)\ fraction by analyzing atomic layer-resolved EELS, to  reveal a charge reconstruction across BTO/STO/LTO layers. As immediately seen, in contrast to conventional STO/LTO interfaces, where  free carriers  density decays   exponentially with the distance from the interface \cite{Nature-Ohtomo-2002,PRL-Cao-2016}, our EELS data reveal the presence of an unusual charge modulation  inside the BTO layers  with  the carrier density comparable or even higher than that of LTO layers (marked as  gray arrows in Fig. 4a). To understand the  unusual charge accumulations observed in the BTO layers we performed first-principles calculations. Figure 4b  shows the atom-resolved orbital occupation for Ti-\textit{d} states atom-by-atom along the \textit{z}-direction defined parallel to the \textit{c}-axis. The density of states in BTO layer exhibits a strong deviation from the exponential decay reported for STO/LTO hetetorstructures \cite{Nature-Ohtomo-2002,PRL-Okamoto-2006,PRL-Cao-2016}, showing consistent feature (blue shadows) with the experimental charge density distribution determined from EELS. More specifically, as seen in Fig. 4b, we find that around the LTO region \textit{d}$_\textrm{xy}$ states are dominantly occupied and the occupation decays exponentially whereas in the BTO region \textit{d}$_\textrm{xz}$/\textit{d}$_\textrm{yz}$ states are mostly occupied with the density shifted toward to the left BTO/STO interface (gray arrows in Fig. 4b). The spatial separation of  \textit{d}$_\textrm{xy}$ and \textit{d}$_\textrm{xz}$/\textit{d}$_\textrm{yz}$ states is  the combined effect of the electrostatic energy and the crystal field splitting; namely, in the LTO region, the electrostatic potential from positively charged (LaO)$^{1+}$ layers dominates and is screened by \textit{d}$_\textrm{xy}$ electrons having in-plane dispersions. In the BTO region, the out-of-plane (or apical) Ti-O distances becomes substantially larger than in-plane Ti-O distances due to the elongated \textit{c}-lattice constant, which in turn lowers the on-site energy of \textit{d}$_\textrm{xz}$/ \textit{d}$_\textrm{yz}$ orbitals and results in the large increase in the  \textit{d}$_\textrm{xz}$/ \textit{d}$_\textrm{yz}$ orbital occupancy compared to STO/LTO heterostructures\cite{PRL-Chang-2013} (see also Supplementary Fig. 5). Around the STO/LTO interface,  \textit{d}$_\textrm{xy}$ states are lower in energy relative to the \textit{d}$_\textrm{xz}$/\textit{d}$_\textrm{yz}$ states from the large modulation in the Ti to out-of-plane (apical) oxygen distance (+0.1/-0.15 \AA) \cite{PRL-Okamoto-2006} and vice versa for the BTO region due to the elongation of \textit{c}-lattice constant increasing average Ti-O distance in the out-of-plane direction. The shift in the charge density can be understood from the polarization gradient of the Ti-O displacements. Around the BTO/STO interface on the left to the BTO region, the negative slope of the Ti-O displacements (black arrows in Fig. 3e) corresponds to the accumulation of the positive ionic charge, functioning as an attractive potential whereas the slope on the right BTO/STO interface (gray arrows in Fig. 3e) is close to zero, resulting in the shift in the charge density to the left BTO/STO interface. In addition, the positive slope in the STO/LTO interface functions as a repulsive potential giving rise to a small shift in $d_\textrm{xy}$ density in the LTO region. Additionally, the  maximum value of Ti-\textit{d} occupancy in the BTO region is  $\sim0.2$ which  exceeds the critical concentration for ferroelectric instability (0.11 per 5 atom unit cell) in bulk BTO \cite{PRL-Wang-2012}. This interesting  observation implies that the charge inhomogeneity and orbital polarization may stabilize the electric  polarization, which requires further theoretical investigation. We note that there are sizable differences in the relative magnitude of the charge density in the LTO and BTO region between the theory and experiment. Since the charge transfer from LTO to STO/BTO region depends mainly on the band alignments between Ti-$d$ bands of LTO and STO/BTO \cite{PRB-2009-Son,PRB-2010-Chen}, the difference may come from the error in the band alignment in the GGA + \textit{U} scheme (see the discussion in the Methods). Another reason for the differences may come from the direct comparison between the integrated local density of Ti $d$-derived states and the Ti valence obtained from EELS. We perform the Bader charge analysis \cite{JPCM-2009-Tang} to check the dependence in calculational methods and find that the difference in the charge around Ti atoms between LTO and BTO region is reduced about four times due to the re-hybridization effect \cite{PRL-2004-Marianetti}, but the characteristic feature of the density profile is maintained for both methods, showing the consistency with respect to the methods used in the comparison.

To summarize, the observed periodic modulation of electric polarization, charge distribution, and orbital occupation  are schematically shown in Fig. 4c. As seen, the polar structure coexists with a 2DEG forming an artificial two-dimensional polar metal at room temperature with the  orbital polarization  (i.e. in-plane \textit{d}$_\textrm{xy}$ state)  controlled by  crystal field engineering. Since coexisting two-dimensional superconductivity and magnetism have been recently reported in electron-doped SrTiO$_3$ \cite{PR-1967-Ko,PRL-2014-Lin,Nature-2013-Richter,NM-2013-Bis,APL-2014-Fuchs}, we can conjecture that  such an artificial non-centrosymmetric 2D metal  may provide a mean to engineer an interesting quantum many-body state with three coexisting phases - ferroelectricity, ferromagnetism, and superconductivity \cite{NP-2017-CR}.

\textbf {\large Methods}

\textbf {Sample synthesis and characterization.}
All films studied in this work were grown on (110)-oriented [orthorhombic notation, corresponding to (001)-orientation in pseudocubic notation] TbScO$_{3}$ substrates (5 $\times$ 5 $\times$ 0.5 mm$^3$) by pulse laser deposition (PLD), using a KrF excimer laser operating at \textit{$\lambda$}~=~248~nm and 2 Hz pulse rate with 2~J/cm$^2$ fluence. The layer-by-layer growth was monitored by \textit{in-situ} reflection-high-energy-electron-diffraction (RHEED). During the growth, the oxygen pressure was kept at $\sim$ 10$^{-6}$ Torr whereas the temperature of the substrates was $\sim$~$850\,^{\circ}{\rm C}$ (from reader of infrared pyrometers). As one of the three components in tricolor superlattice (BaTiO$_3$)$_{10}$/(SrTiO$_3$)$_3$/(LaTiO$_3$)$_3$, bulk BaTi$^{4+}$O$_3$ with 3\textit{d}$^{0}$ electron configuration, a well-known ferroelectric material, undergoes complex structural and ferroelectric phase transitions on cooling, e.g., from cubic to tetragonal near 400~K, tetragonal to orthorhombic near 280 K, and orthorhombic to rhombohedral near 210~K (ferroelectric properties are present in the later three phases) \cite{PRL-1997-Ish,JPCM-2002-Hay,ARMR-2007-Schlom}; bulk SrTi$^{4+}$O$_3$ with 3\textit{d}$^{0}$ electron configuration is a band insulator (gap size $\sim$ 3.3 eV) \cite{PRB-2009-Taki}; bulk LaTi$^{3+}$O$_3$ with 3\textit{d}$^{1}$ electron configuration is a Mott insulator and undergoes G-type antiferromagnetic (AFM) transition below 146~K \cite{PRB-2003-Cwik}. In bulk, the lattice parameters are \textit{a} = 3.905 \AA~for cubic SrTiO$_3$; \textit{a} = 4.00 \AA~for cubic BaTiO$_3$; \textit{a} = 5.466, \textit{b} = 5.727, \textit{c} = 7.915, $\sqrt{\textit{a}^2+\textit{b}^2}$/2 = 3.9584, \textit{c}/2 = 3.9575 \AA~for orthorhombic TbScO$_{3}$; \textit{a} = 5.595, \textit{b} = 5.604, \textit{c} = 7.906, $\sqrt{\textit{a}^2+\textit{b}^2}$/2 = 3.959, \textit{c}/2 = 3.953 \AA~for orthorhombic LaTiO$_{3}$. Therefore, the SrTiO$_3$ layers of BaTiO$_3$/SrTiO$_3$/LaTiO$_3$ superlattice on TbScO$_{3}$ substrate are under tensile strain, whereas the BaTiO$_3$ layers are under compressive strain. To guarantee the intriguing electronic states arise from the interfaces of (BaTiO$_3$)$_{10}$/(SrTiO$_3$)$_3$/(LaTiO$_3$)$_3$ and are not induced by the BaTiO$_3$, SrTiO$_3$, and LaTiO$_3$ layers themselves due to the oxygen doping (LaTiO$_{3+\textit{$\delta$}}$) or vacancies (SrTiO$_{3-\textit{$\delta$}}$ and BaTiO$_{3-\textit{$\delta$}}$), the right Ti valence states in single BaTiO$_3$, SrTiO$_3$, and LaTiO$_3$ films were confirmed by XAS and temperature-dependent electrical transport. The structural quality of the superlattices is further confirmed by XRD using Cu K$_{\textrm{$\alpha$}1}$ radiation. Electrical transport properties of all films were measured in Van der Pauw geometry by a Physical Properties Measurement System (PPMS, Quantum design) operating in high-resolution d.c. mode. The sheet carrier density (\textit{n}$_\textrm{s}$) is estimated from \textit{n}$_\textrm{s}$ = 1/($eR$$_\textit{H}$), where the Hall coefficient \textit{R}$_\textit{H}$ is calculated with \textit{R}$_\textit{H}$ = \textit{V}$_\textit{H}$/\textit{I}\textit{B} (\textit{V}$_\textit{H}$ is the Hall voltage, \textit{I} is the driven current, and B is the out-of-plane magnetic field). As seen in Fig. 1c, there are 4 metallic STO/LTO interfaces in the superlattice BTO/STO/LTO. Therefore, the carrier density per interface is \textit{n} = \textit{n}$_\textrm{s}$/4.

\textbf{STEM/EELS measurement.}
Cross-sectional TEM samples are prepared by Focused Ion Beam(FIB) with Ga$^{+}$ ions followed with Ar$^{+}$ ion milling \cite{PRB-Wang-2016}. STEM/EELS experiments are performed on a double aberration corrected microscope JEOL-ARM200F operating at 200 keV, with a Dual energy-loss spectrometer. The structural features of the films are studied by atomic resolution HAADF- and ABF-STEM imaging. Collection angles for HAADF- and ABF-STEM imaging are 67 to 375 mrad and 11 to 23 mrad, respectively, and the convergence semi-angle is about 21 mrad. The atomic positions are obtained using two-dimensional Gaussian fitting following the maximum intensity. When collecting EELS spectra, the microscope conditions are optimized for EELS collection with a probe size of $\sim$ 0.9 \AA~and a convergence semi-angle about  20 mrad. Line-scanning EELS spectra across the BTO/STO/LTO interfaces are acquired with a step size of 0.12 \AA, and a dwell time of 0.05 s per pixel. The energy resolution is about 0.7-1 eV depending on the energy dispersion selected. To measure the elemental concentration of all the cations, a lower dispersion is selected to simultaneously collect Ti L$_{2,3}$-, O K-, Ba M$_{4,5}$-, La-M$_{4,5}$, Sc L$_{2,3}$-, and Tb M$_{4,5}$-edges. To extract the bonding information, a higher energy dispersion is chosen for fine structure analysis of Ti L$_{2,3}$-edge. Dual EELS mode is used to simultaneously acquire both the zero-loss and core-loss EELS spectra to ensure that the intrinsic energy drift during spectrum acquisition can be compensated. The background of EELS spectra is subtracted with power-law function and multiple scattering was corrected by deconvolution. The EELS elemental profile is obtained by integrating the Ti L$_{2,3}$-, O K-, Ba M$_{4,5}$-, La-M$_{4,5}$, Sc L$_{2,3}$-, and Tb M$_{4,5}$-edges, respectively.

\textbf {SHG measurement.}
SHG measurements are performed in a far-field transmission geometry using an 800 nm fundamental laser beam generated by a Spectra-Physics Solstice Ace Ti:Sapphire femtosecond laser system ($<$100 fs, 1 kHz). The schematic of experimental setup is shown in Fig.~2a, where a linear polarized fundamental 800 nm beam with polarization direction $\varphi$ is incident on the sample at a tilted angle $\theta$, the transmission SHG field is first spectrally filtered, then decomposed into \textrm{s}-/\textrm{p}-polarized components (\textbf{E}$_{2\textit{$\omega$},\textrm{s}}/\textbf{E}_{2\textit{$\omega$},\textrm{p}}$) and finally detected by a photon multiplier tube. The in-plane orientation of the sample is controlled by angle \textit{$\beta$}. Systematic polarimetry is done by scanning polarization direction \textit{$\varphi$} with fixed \textit{$\theta$} and \textit{$\beta$}. 

Theoretical modeling of SHG data is described below. The fundamental field with polarization direction \textit{$\varphi$} inside the sample can be written as
\begin{eqnarray}
{\textbf{E}_{\omega,1}'}&=&[\cos⁡(\theta^\prime)  \cos⁡(\beta) \cos⁡(\varphi) t_\textrm{p} -\sin⁡(\beta) \sin⁡(\varphi) t_\textrm{s}]E_\omega\\
{\textbf{E}_{\omega,2}'}&=&[\cos⁡(\theta^\prime)  \sin⁡(\beta) \cos⁡(\varphi) t_\textrm{p} +\cos⁡(\beta) \sin⁡(\varphi) t_\textrm{s}]E_\omega\\
{\textbf{E}_{\omega,3}'}&=&-\sin⁡(\theta^\prime)  \cos⁡(\varphi) t_\textrm{p} E_\omega,
\end{eqnarray}

where $\sin(\theta^\prime)=\sin(\theta)/n$, $n$ is the refractive index, $t_\textrm{p} = 2\cos(\theta)/[n\cos(\theta)+\cos(\theta^\prime)]$ and  $t_\textrm{s}=2\cos(\theta)/[\cos(\theta)+n\cos(\theta^\prime)]$ are Fresnel coefficients, $E_\omega$ is the magnitude of electric component in air. Under Voigt notation, the SHG field generated through nonlinear optical process can be expressed as \textbf{E}$_{2\omega,i}' = \textbf{d}_{ij}\textbf{E}_{\omega,j}'^{Voigt}$, where normalized SHG \textbf{d} matrix for 4$mm$ and $mm$2 symmetry is
\begin{eqnarray}
        \textbf{d}^{4mm} =\left( 
        \begin{array}{cccccc}
        0 & 0 & 0 & ~0 & ~~\textbf{d}_{15} & ~~0 \\
        0 & 0 & 0 & ~\textbf{d}_{15} & ~~0 & ~~0 \\
        \textbf{d}_{31} & \textbf{d}_{31} & \textbf{d}_{33} & ~0 & ~~0 & ~~0 
        \end{array}
        \right),\\
        \nonumber\\
        \textbf{d}^{mm2} =\left( 
        \begin{array}{cccccc}
        0 & 0 & 0 & ~0 & ~~\textbf{d}_{15} & ~~0 \\
        0 & 0 & 0 & ~\textbf{d}_{24} & ~~0 & ~~0 \\
        \textbf{d}_{31} & \textbf{d}_{32} & \textbf{d}_{33} & ~0 & ~~0 & ~~0 
        \end{array}
        \right)
\end{eqnarray}

In practice, all the coefficients in \textbf{d} tensor are normalized to \textbf{d}$_{15}$. By ignoring the index dispersion, that is, $n=n_\omega\approx n_{2\omega}$. The transmitted SHG field is
\begin{eqnarray}
        \textbf{E}_{2\omega,\textrm{p}}&=&[\cos⁡(\theta^\prime) \cos⁡(\beta) \textbf{E}_{2\omega,1}'+\cos⁡(\theta^\prime)\sin⁡(\beta)\textbf{E}_{2\omega,2}' -\sin⁡(\theta^\prime) \textbf{E}_{2\omega,3}' ] t_\textrm{p}^\prime\\
        \textbf{E}_{2\omega,\textrm{s}}&=&[-\sin(\beta)\textbf{E}_{2\omega,1}'+\cos(\beta)\textbf{E}_{2\omega,2}']t_\textrm{s}^\prime
\end{eqnarray}
where $t_\textrm{p}^\prime = 2n\cos(\theta^\prime)/[n\cos(\theta)+\cos(\theta^\prime)]$,   $t_\textrm{s}^\prime=2n\cos(\theta^\prime)/[\cos(\theta)+n\cos(\theta^\prime)]$, the subscripts $s$/$p$ denote $s$/$p$ components of corresponding field variables. Finally, we have below equations for SHG measured intensity
\begin{eqnarray}
{\textbf{I}_{2\omega,\textrm{s}}}&=&\alpha|\textbf{E}_{2\omega,\textrm{s}}|^2   \\
{\textbf{I}_{2\omega,\textrm{p}}}&=&\alpha|\textbf{E}_{2\omega,\textrm{p}}|^2
\end{eqnarray}
where \textit{$\alpha$} is scaling factor. 

\textbf{First-Principles calculations.}
We performed first-principles density-functional-theory calculations with the generalized gradient approximation plus \textit{U} (GGA + \textit{U}) method using the Vienna {\it ab-initio} simulation package \cite{PRB-1996-Kre,PRB-1999-Kre}. The Perdew-Becke-Erzenhof parametrization \cite{PRL-1996-Perdew} for the exchange-correlation functional and the rotationally invariant form of the on-site Coulomb interaction \cite{PRB-1995-Lie} are used with \textit{U} = 3 and \textit{J} = 0.68 eV for the titanium \textit{d} states \cite{EPL-2008-Zhong,EPL-2011-Okatov,Science-2011-Jang} and \textit{U}$_\textrm{f}$ = 11 and \textit{J}$_\textrm{f}$ = 0.68 eV for the lanthanum $f$ bands away from the Fermi energy \cite{PRB-1994-Czy}. Our choice of \textit{U} value reproduces the correct ground states of bulk LaTiO$_{3}$ and LaTiO$_{3}$/SrTiO$_{3}$ and LaAlO$_{3}$/SrTiO$_{3}$ superlattices\cite{EPL-2011-Okatov,EPL-2008-Zhong,PRL-2007-Pentcheva} while maintaining reasonable agreement in the bulk band alignments between filled Ti-\textit{d} derived lower Hubbard band of LaTiO$_{3}$ and empty Ti-\textit{d} derived states of SrTiO$_{3}$ and BaTiO$_{3}$ compared with those from the hybrid functional\cite{JPC-2003-Heyd,PRB-2008-Wahl,PRB-2012-He}. The projector augmented wave method \cite{PRB-1994-Blo} are used with pseudopotentials contain 6 valence electrons for O ($2s^{2}2p^{4}$), 12 for Ti ($3s^{2}3p^{6}3d^{2}4s^{2}$), 10 for Sr ($4s^{2}4p^{6}5s^{2}$), 10 for Ba ($5s^{2}5p^{6}6s^{2}$), and 11 for La ($5s^{2}5p^{6}5d^{1}6s^{2}$). We used an energy cutoff of 500 eV and $k$-point sampling on a $8\times 8\times 1$ grid with $1\times 1$ in-plane unit cell. Although our choice of the unit cell does not include the GdFeO$_{3}$ type of octahedral rotation and tilts observed in bulk LaTiO$_{3}$\cite{PRL-2003-MM,PRL-2004-PE}, we find that the orbital polarization of Ti-$d$ states for (LaTiO$_{3}$)$_{2}$/(SrTiO$_{3}$)$_{4}$ superlattices are similar in the magnitude regardless of the presence of the octahedral rotation and tilts for the interface and SrTiO$_{3}$ region with some deviations in the LaTiO$_{3}$ region. Since our main focus is the properties of the interface electron gas, it is reasonable to use of the $1 \times 1$ in-plane unit for the large unit-cell size of the tricolor superlattices. The atomic positions and $c$-lattice constants are fully relaxed with 0.02 eV per \AA ~force threshold  while $a$ and $b$ lattice constants are fixed to experimental in-plane lattice constants of TbScO$_3$. The layer resolved charge distribution in Fig. 4b and Supplementary Fig. 5a,b are calculated by integrating the Ti-$d$ projected density of states in the energy windows from -1 eV to the Fermi energy (Supplementary Fig. 5c) and is normalized to satisfy that the total number of electron is the same as the number of LaO layers.

\textbf{Data availability.}
The data and code that support the findings of this study are available from the corresponding authors upon reasonable request.

\textbf {\large Acknowledgments}

The authors acknowledge insightful discussions with S. W. Cheong, S. J. Lim, E. W. Plummer, Jiandi Zhang, and Hangwen Guo. J.C. is supported by the Gordon and Betty Moore Foundation EPiQS Initiative through Grant No. GBMF4534. Y.C. is supported by the Pioneer Hundred Talents Program of Chinese Academy of Sciences. Z.W. is supported by US Department of Energy (DOE) under Grant No. DOE DE-SC0002136. The electronic microscopic work done at Brookhaven National Laboratory is sponsored by the US DOE Basic Energy Sciences, Materials Sciences and Engineering Division under Contract DE-AC02-98CH10886. M.K. and X.L. were supported by  the DOE/BES, under award number DE-SC0012375 for their synchrotron work at  ALS. The work at Pennsylvania State University is supported by the DOE/BES, under award number DE-SC0012375 (Y.Y., S.N., H.A., and V.G.). The XMaS beamline at the ESRF is a mid-range facility supported by UK EPSRC. The Advanced Light Source is supported by the Director, Office of Science, Office of Basic Energy Sciences, of the U.S. Department of Energy under Contract No. DE-AC02-05CH11231. The first-principles calculations were performed on the Rutgers University Parallel Computer (RUPC) cluster and supported by the Office of Naval Research Grant No. N00014-14-1-0613.
\\

\textbf {\large Author contributions}

Y. C. and J. C. conceived the project. Y. C., M. K., X. L., P. S., A. N., E. A., and J. C. synthesized and characterized the samples. Z. W. and Y. Z. carried out STEM/EELS. S. P. and K. R. performed the calculations. Y. Y., S. N., H. A. and V. G. measured SHG. P. T. and P. R. collected low temperature XRD data.  All authors discussed the results.

\textbf {\large Additional information}

\textbf{Supplementary Information} is available in the online version of paper. Reprints and permissions information is available online at www.nature.com/reprints. Correspondence and requests for materials should be addressed to Y.C.

\textbf{Competing interests:} The authors declare no competing financial or non-financial interests.

\end{document}